# Reciprocal Symmetry and Classical Discrete Oscillator Incorporating Hal-Integral Energy Levels

Mushfiq Ahmad

Department of Physics, Rajshahi University, Rajshahi, Bangladesh

E-mail: mushfiqahmad@ru.ac.bd

**Abstract**

Classical oscillator differential equation is replaced by the corresponding (finite time) difference equation. The equation is, then, symmetrized so that it remains invariant under the change $d \rightarrow -d$, where $d$ is the smallest span of time. This symmetric equation has solutions, which come in reciprocally related pairs. One member of a pair agrees with the classical solution and the other is an oscillating solution and does not converge to a limit as $d \rightarrow 0$. This solution contributes to oscillator energy a term which is a multiple of half-integers.

## 1. Introduction

The differential equation

$$\frac{df}{dt} = iwf \qquad (1.1)$$

has a unique solution. The corresponding finite difference equation has more solutions[1]. When the function represents a harmonic oscillator, different solutions will contribute to oscillator energy in different ways. We intend to study these contributions and compare them to the corresponding quantum mechanical values.

## 2. Oscillator Finite Difference Equation

Classical simple harmonic oscillator function $f$ (with angular speed $w$) satisfies differential equation (1.1)

To exploit its symmetry properties we replace the above differential equation by the corresponding symmetric finite difference equation[2]

$$\frac{Dg_{\pm}}{D(t,\delta)} = iW \cdot g_{\pm} \qquad (2.1)$$

where

$$\frac{Dg_{\pm}(w,t)}{D(t,\delta)} = \frac{g_{\pm}(w,t+\delta) - g_{\pm}(w,t-\delta)}{2\delta} \qquad (2.2)$$

The above difference quotient has the following symmetry under the change $\delta \to -\delta$

$$\frac{Dg_{\pm}}{D(t,-\delta)} = \frac{Dg_{\pm}}{D(t,\delta)} \qquad (2.3)$$

We require that at least one of the solutions, $g_+$, of (2.1) should go over to (1.1) in the limit $\delta \to 0$

$$\frac{Dg_+}{D(t,\delta)} = iW(w)g_+ \xrightarrow{\delta \to 0} \frac{df}{dt} = iwf \qquad (2.4)$$

With

$$g_+ \xrightarrow{\delta \to 0} f \text{ and } W \xrightarrow{\delta \to 0} w \qquad (2.5)$$

### 3. Reciprocal Symmetry

Let $g_{\pm}$ be of the form $g_{\pm} = (\pm a)^{\pm t/\delta}$ so that

$$g_+(w,t) = (-1)^{t/\delta} g_-(w,t) \qquad (3.1)$$

Consider equation (2.1)

$$\frac{g_+(w,t+\delta) - g_+(w,t-\delta)}{2\delta} = iW \cdot g_+(w,t) \qquad (3.2)$$

Using (3.1) we find that $g_-$ also satisfies the equation. This establishes reciprocal symmetry of (2.1), that the equation remains invariant under transformation (3.1).

### 4. Reciprocal symmetric Solutions

(2.1) has a pair of solutions

$$g_\pm(w\delta/2) = \left( \pm \frac{1 \pm i\sin(\frac{w\delta}{2})}{1 \mp i\sin(\frac{w\delta}{2})} \right)^{t/\delta} = (\pm 1)^{t/\delta} \exp(\pm iwt) \tag{4.1}$$

$g_+$ and $g_-$ satisfy (2.1) with

$$W = \frac{\sin(w\delta)}{\delta} \tag{4.2}$$

We may write

$$g_+ = \exp\left(\frac{(2n)\pi t}{\delta} i\right) \exp(iwt) = \exp(iw_+ .t) \tag{4.3}$$

$$g_- = \exp\left(\frac{(2n+1)\pi t}{\delta} i\right) \exp(-iwt) = \exp(iw_- .t) \tag{4.4}$$

where

$$w_+ = (2n\pi)/\delta + w = y_+ + w \tag{4.5}$$

$$w_- = (2n+1)\pi/\delta - w = y_- - w \tag{4.6}$$

## 5. Classical and Half-Integral Energy Levels

The energy of the oscillator is proportional to

$$(w_+)^2 = y_+^2 + 2y_+ w + w^2 = \{(2n\pi)/\delta\}^2 + 2\{(2n\pi)/\delta\}w + w^2 \tag{5.1}$$

$$(w_-)^2 = y_+^2 + 2y_- w + w^2 = \{(2n+1)\pi/\delta\}^2 + 4\{(n+1/2)\pi/\delta\}w + w^2 \tag{5.2}$$

For $n=0$ (5.1) gives the classical value. The middle term of (5.2) is a product of half-integers and $w$. To this extent it corresponds to quantum mechanical value.

## 4. Conclusion

We have replaced oscillator differential equation by the corresponding symmetric discrete equation (2.1). This has brought to surface important parts of oscillator function, which were lost in the conventional solution. These parts contain discrete – integral and half integral -- energy levels.